\documentclass[prb,showpacs,twocolumn,floatfix,superscriptaddress]{revtex4}
\usepackage{epsfig}
\usepackage{amsmath,amssymb,mathrsfs}
\newcommand\varH{\mathscr{H}}%
\newcommand{\beq}{\begin{equation}}
\newcommand{\eeq}{\end{equation}}
\newcommand{\bea}{\begin{eqnarray}}
\newcommand{\eea}{\end{eqnarray}}

\begin{document}

\title{Rashba interaction in quantum wires with in-plane magnetic fields}
\author{Lloren\c{c} Serra}
\affiliation{Departament de F\'{\i}sica, Universitat de les Illes Balears, E-07122 
Palma de Mallorca, Spain}
\affiliation{Institut Mediterrani d'Estudis Avan\c{c}ats IMEDEA (CSIC-UIB), 
E-07122 Palma de Mallorca, Spain}
\author{David S\'anchez}
\affiliation{Departament de F\'{\i}sica, Universitat de les Illes Balears, E-07122 
Palma de Mallorca, Spain}
\author{Rosa L\'opez}
\affiliation{Departament de F\'{\i}sica, Universitat de les Illes Balears, E-07122 
Palma de Mallorca, Spain}

\begin{abstract}
We analyze the spectral and transport properties of ballistic
quasi one-dimensional systems in the presence of spin-orbit coupling and
in-plane magnetic fields. Our results demonstrate that Rashba precession 
and intersubband coupling must be treated on equal footing
for wavevectors near the magnetic field induced gaps.
We find that intersubband coupling limits the occurrence of negative effective
masses at the gap edges and modifies the linear conductance curves
in the strong coupling limit. 
The effect of the magnetic field on the spin textured orientation
of the wire magnetization is discussed.
\end{abstract}

\pacs{71.70.Ej, 72.25.Dc, 73.63.Nm}
\maketitle

\section{Introduction} 
Controllable manipulation of electron spins
with electric fields (gates) is a central requirement to
spintronic devices. Semiconductor heterostructures
offer the possibility of electric control of spins through
intrinsic {\em spin-orbit} interactions. A major contribution
to spin-orbit effects in two-dimensional
(2D) electron gases of narrow-gap semiconductors
originates from the macroscopic electric field confining
the electron gas.\cite{Ras60}
This implies an asymmetry in the quantum-well potential
(the Rashba effect), which can be further tuned with gate voltages.
The resulting spin-orbit coupling has been demonstrated in 
experiments.\cite{Nit97,Eng97,Scha04}

In this paper, we are concerned with the Rashba interaction in
ballistic {\em quantum wires}. Interestingly,
these systems have been proposed as basic elements for
practical applications such as spin-dependent
field-effect transistors\cite{Dat90} and spin filters\cite{stre03}
and have been also considered as detectors of entangled pairs of
electrons\cite{egu02} and of the hyperfine nuclear spin  
dynamics.\cite{nes04}
The Rashba interaction is described
by the Hamiltonian,
\begin{equation}\label{eq_hras}
{\varH}_R=-\frac{\alpha_1}{\hbar}p_x\sigma_y
+\frac{\alpha_2}{\hbar}p_y\sigma_x\,,
\end{equation}
where $\vec p=(p_x,p_y)$ is the linear momentum
and $\sigma_x$, $\sigma_y$ are Pauli matrices.
Spin transistors exploit
the Rashba-induced precession of spins ($\alpha_1$ term for transport 
along the $x$-direction). However, in Eq.~(\ref{eq_hras}) there is 
an additional term, proportional to $\alpha_2$, which mixes nearest 
subbands with opposite spins and induces anticrossings at the 
degeneracy points of the wire's
energy spectrum.\cite{Mor99,Mir01}
Obviously,
in a device one has $\alpha_1=\alpha_2$,
but it is of conceptual interest to distinguish in Eq.~(\ref{eq_hras})
between the two contributions (see below). 

When an in-plane magnetic field is externally applied or it arises
from ferromagnetic leads attached to the wire, the Zeeman splitting
opens gaps in the wire spectrum,\cite{stre03,Yur04,cah04,nes04,per05}
strongly affecting the transport properties.
It also influences the spin dynamics since
whereas the Rashba precession randomizes the spin direction
(a common spin quantization axis can be defined only for large wavevectors),
the magnetic field tends to align the spins parallel to it.
Many works neglect\cite{Yur04,cah04,nes04,per05}
the effect of Rashba intersubband coupling (RIC)
in quantum wires in the presence of in-plane magnetic fields.
Here we demonstrate that both Rashba precession and intersubband
coupling must be treated {\em on equal footing}
for wavevectors near the subband gaps.
We find that RIC hinders the formation of subband maxima, smoothing the gap 
edges and, as a consequence, strongly affecting the conductance steps.
This conclusion is most relevant to the strong coupling limit, when 
spin-orbit and confinement energy scales are of the 
same order of magnitude.
Section II presents the model and analyzes its characteristic energy subbands.
In Sec.\ III we show the spin expectation values and magnetization
distributions while Sec.\ IV focusses on the linear conductance and,
finally, the conclusions are contained in Sec.\ V.

\section{Hamiltonian and energy spectrum}

\begin{figure}[b]
\centerline{\psfig{figure=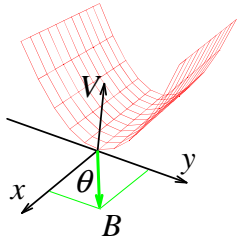,width=1.25in,clip=}}
\caption{(Color online) Schematic representation of the 
wire potential $\varH_{\rm conf}\equiv V(y)=m\omega_0^2 y^2/2$
and magnetic field orientation considered in this work.} 
\label{fig1}
\end{figure}

We consider a quantum wire, formed when
a 2D gas is further confined in one direction [see Fig.\ \ref{fig1}]. 
The confinement 
is assumed parabolic in the $y$-direction, 
$\varH_{\rm conf}=m\omega_0^2 y^2/2$, giving a wire orientation
along $x$. An in-plane magnetic field 
$\vec{B}=(B\cos\theta,B\sin\theta)$ acts through the Zeeman Hamiltonian
$\varH_Z=g \mu_B B (\cos\theta\sigma_x+\sin\theta\sigma_y)/2$.
Orbital magnetic effects are absent in this geometry since they 
arise only from perpendicular fields.\cite{Scha04,Byc90,Deb04} 
Adding all contributions, the resulting quasi-1D Hamiltonian reads,
\begin{equation}\label{totham}
\varH=(p_x^2+p_y^2)/2m+\varH_{\rm conf}+\varH_Z+\varH_R\; .
\end{equation}

Since the system is translationally invariant along $x$, the wave 
function dependence on this variable is $e^{ikx}$, with $k$ 
the propagation wavevector.
It is then useful to recast the Hamiltonian
for a given wavevector, $\varH_k$, in dimensionless form:\cite{Deb04}
\begin{eqnarray}\label{secondquantization}
\frac{\varH_k}{\hbar\omega_0}&=&\left(\hat{n}_k+\frac{1}{2}\right)
+\frac{1}{2}\left(\frac{l_0}{l_Z}\right)^2\left(\cos\theta\sigma_x 
+\sin\theta\sigma_y \right) \nonumber\\
&+&\!\!\frac{(k l_0)^2}{2}-\!\frac{l_0}{2l_1}
    (kl_0)\sigma_y \!+\! \frac{il_0}{2\sqrt{2}\,l_2}
    \!\left( \hat{a}_k^\dagger-\hat{a}_k \right)\!\sigma_x\,,
\end{eqnarray}
with the characteristic lengths $l_0=\sqrt{\hbar/m\omega_0}$ (confinement),
$l_Z=\sqrt{\hbar^2/mg \mu_B B}$ (Zeeman coupling, $g>0$)\cite{notefoot}
and $l_i=\hbar^2/2 m \alpha_i$ ($i\in 1,2$, Rashba interaction).
We use in Eq.~(\ref{secondquantization}) the bosonic operator
$a_k$ ($a_k^\dagger$) which lowers (raises) a subband index
for fixed $k$, thereby $\hat{n}_k=a_k^\dagger a_k$ is the number
operator. The third term in the right-hand side of
Eq.~(\ref{secondquantization}) describes the free propagation motion
in the $x$-direction whereas the forth
and fifth contributions correspond to Rashba precession ($l_1$) and
to RIC ($l_2$), respectively. The $l_1$ Rashba coupling induces subband
spin splitting whereas it is clear from Eq.~(\ref{secondquantization})
that the $l_2$ term couples adjacent subbands with opposite spins.
In writing $\varH_k$ we have omitted the Coulomb interaction between
the electrons since its effect can be taken into account, at least in part,
via a renormalized Rashba coefficient.\cite{Che99} 

\begin{figure}[t]
\centerline{\psfig{figure=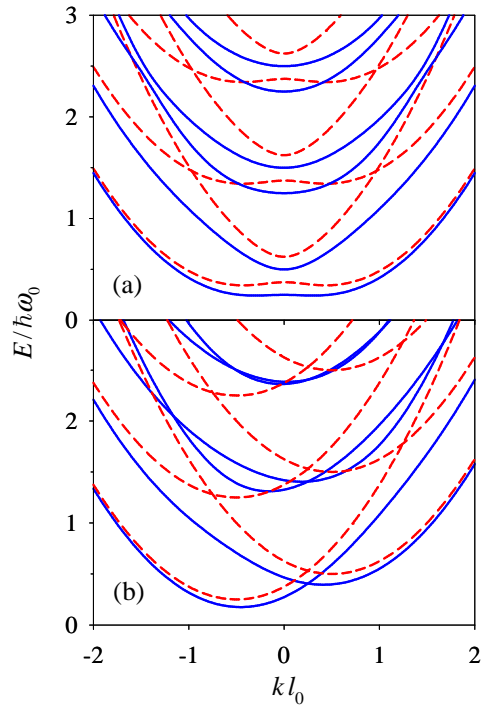,width=2.5in,clip=}}
\caption{(Color online) 
(a) Dispersion relation for $\theta=0$, $l_Z=2l_0$, and
$l_1=l_0$. Solid line corresponds to the case where the
Rashba intersubband coupling is fully included $(l_{2}=l_{1})$ while
dashed line shows the case without it ($l_{2}=\infty$).
(b) Same as (a) for $\theta=\pi/2$.} 
\label{fig2}
\end{figure}

Due to the presence of RIC, $\varH_k$
is not diagonalizable except for a special wavevector (see below).
Yet, it is easy to show that for fields parallel to the wire
the two Rashba terms contribute equally.
Without Rashba interaction the energy dispersion of a
given subband is
$E_{n k \eta}^{(0)}=(n+1/2)+(k l_0)^2/2+\eta (l_0/l_Z)^2/2$ with 
$n=0,1,2,\ldots$ and $\eta=\pm 1$.
We find the energy correction to first order in the Rashba couplings:
\begin{equation}\label{firstorder}
E_{n k \eta}^{(1)}=-\eta\frac{1}{2}(k l_0)\frac{l_0}{l_1}\sin\theta\,.
\end{equation}
Equation~(\ref{firstorder}) already shows two noticeable features:
(i)  When $\theta\ne 0$ there is a Rashba-induced splitting proportional
to $k$, well known from studies of spin-orbit effects at $B=0$
(see, e.g., Ref.\ \onlinecite{Val03}), which combines with the Zeeman
splitting to yield multiple subband crossings; and (ii)
RIC (the $l_2$ term) does not contribute to first order but for $\theta=0$ 
the Rashba precession ($l_1$) term is also zero. This implies that when $\theta=0$
both Rashba terms contribute to second order in the couplings.
The full second-order correction reads:
\begin{eqnarray}\label{secondorder}
E_{n k \eta}^{(2)}&=&\eta\frac{l_0^2}{4l_1^2}(k l_0)^2
\frac{l_Z^2}{l_0^2} \cos^2\theta \nonumber \\
&-&\frac{l_0^2}{8l_2^2} \left[ 1+ \sin^2\theta
\frac{(l_0/l_Z)^4+\eta(2n+1)(l_0/l_Z)^2}{1-(l_0/l_Z)^4} \right]
\;,\nonumber\\
\end{eqnarray}
Adding all corrections one has
$E_{n k \eta}\simeq E_{n k \eta}^{(0)}+E_{n k \eta}^{(1)}+E_{n k \eta}^{(2)}$,
which is valid for $l_0,l_Z\ll l_1,l_2$ and $l_0\neq l_Z$. 
It should be emphasized that in the above perturbative 
analysis we have taken as expansion parameters
$l_0/l_1$ and $l_0/l_2$, assuming that the remaining factors of the
two Rashba terms are similar, i.e., that $kl_0\simeq 1$. Obviously,
in the limit
of very large $k$ the Rashba precession term, proportional
to $kl_0$, will clearly dominate over the RIC. 

Further progress is made if we diagonalize the total Hamiltonian
in the absence of RIC, i.e., setting $l_2\to\infty$
in Eq.~(\ref{secondquantization}). The exact eigenstates
and eigenenergies (in $\hbar\omega_0$ units),
\begin{eqnarray}\label{states_s1}
\!\!\!\!\psi_{nk \eta}(x,y) &=&\frac{e^{ikx}}{\sqrt{2}}\, \phi_{n}(y)
\!\!\left[ e^{i\Omega_{k}/2}|\uparrow\rangle + \eta
  e^{-i\Omega_{k}/2}|\downarrow\rangle\right],\\
E_{nk\eta} &=& \left(n+\frac{1}{2}\right) + \frac{1}{2} (kl_0)^2
+\eta|z_{k}|\label{energies_s1}
\;,
\end{eqnarray}
are labeled, as before, with three quantum numbers: the propagation
wavevector $k$, the index $n$
of the corresponding 1D harmonic oscillator wavefunction $\phi_n(y)$,
and the branch-splitting quantum index $\eta=\pm1$. 
The notation has been simplified by defining the complex number
$z_{k}=(l_0/l_Z)^2 e^{-i\theta}/2+ i (k l_0/2)(l_0/l_1)$
and its argument $\Omega_{k}={\rm Arg}[z_k]$. This complex quantity
is taking into account the relative importance of Zeeman and Rashba 
precession energies, as can be seen more clearly when rewriting it like
$z_k=[(g \mu_B B/2)e^{-i\theta} + i \alpha_1 k]/\hbar\omega_0$.

The energy spectrum of Eq.\ (\ref{energies_s1}), for $\theta=0$ and 
$\theta=\pi/2$ is plotted with dashed lines in
Figs.~\ref{fig2}(a) and (b) for $l_Z=2l_0$ and strong spin-orbit
$l_1=l_0$ (taking $\alpha\approx 10$ meV$\,$nm and InAs parameters
the above values would correspond to $\hbar\omega_0\approx 0.2$ meV and 
$B\approx 0.1$ T, which can be achieved experimentally).\cite{Scha04}
For both angles the spectrum
shows energy crossings between subbands with $n\neq n'$
and opposite $\eta$. For $\theta=\pi/2$ there are  
additional crossing points for $kl_0\approx0.2$ between 
subbands of opposite $\eta$ and equal $n$.
Noticeably, for $\theta=0$ the Zeeman field produces gaps at $k=0$
and around that point the spectrum shows local maxima for the $\eta=-1$ subbands. From Eq.~(\ref{energies_s1}) we see
that a local maximum (minimum) occurs at the $\eta=-1$ subbands
when $l_Z>\sqrt{2}l_{1}$ ($l_Z<\sqrt{2}l_{1}$).
Thus, changing the magnetic field affects dramatically 
the spectrum and, as discussed below, the transport properties.

Solid lines in Figs.~\ref{fig2}(a) and (b) show the effect
of RIC. We have obtained the spectrum with a direct
numerical diagonalization of Eq.~(\ref{secondquantization})
in the basis of Eq.\ (\ref{states_s1}), truncating to a large enough $n$. 
The intersubband matrix elements 
of $\varH_{\alpha_2}\equiv(il_0/2\sqrt{2}l_2)(a^\dagger-a)\sigma_x$ read
\begin{eqnarray}
\label{matrixelement1}
\langle n k \eta | \varH_{\alpha_2} | n' k' \eta' \rangle &=&
\frac{i}{4\sqrt{2}}
\frac{l_0}{l_2}\delta_{k,k'}  
\left( \eta e^{i\Omega_k}+\eta' e^{-i\Omega_k}\right)\nonumber\\
&&\!\!\!\!\!\!\!\!\!\!\!\!\times
\left(\delta_{n',n-1}\sqrt{n}-\delta_{n',n+1}\sqrt{n+1}\,\right)\,.
\end{eqnarray}
As an alternative method, we have also discretized
$\varH_k$ in real space with finite differences and 
diagonalized the resulting matrix.
Both methods give identical results. When RIC
is fully included we find that the crossings of different-$n$ subbands 
(dashed lines) become anticrossings (solid lines) as expected.
Importantly, for $\theta=0$ and high energy subbands RIC converts the maxima into minima
[see, e.g., the second and third subbands in Fig~\ref{fig2}(a)].

Figure \ref{fig2} also shows a conspicuous downward shift of the solid lines
with respect to the dashed ones for $k$-values close to the central gaps 
(for $\theta=0$) or degeneracy points (for $\theta=\pi/2$).
This can be easily explained by noting that
the minimum gap condition for $\theta\ne\pi/2$ or
crossing point for $\theta=\pi/2$ is given by ${\rm Im}[z_{k_g}]=0$ 
with solution $k_g=(l_1/l_Z^2)\sin\theta$.
When this occurs the Hamiltonian~(\ref{secondquantization})
can be {\em exactly} solved since $\varH_{\alpha_2}$  does not couple states
with different $\eta$-indices [see Eq.~(\ref{matrixelement1})
and note that $\eta+\eta'=2\eta\delta_{\eta\eta'}$].
The Hamiltonian for each $\eta$,
\begin{eqnarray}\label{eq_exact}
\varH_\eta &=& \frac{\hbar^2 k_g^2}{2m}+
\frac{(p_y+\eta m\alpha_2/\hbar)^2}{2m}
+\frac{1}{2} m \omega_0 y^2\nonumber\\
&+&
\eta|z_{k_g}|
-\frac{\alpha_2^2m }{2\hbar^2}
\,,
\end{eqnarray}
represents a harmonic oscillator with a shifted transverse momentum and a 
global negative energy constant. 
As a result, the exact states at the minimum gap are those of Eq.\ (\ref{states_s1})  
shifted in momentum, $e^{-i\eta{m}\alpha_2 y/\hbar^2}\psi_{nk_g \eta}(x,y)$, and 
the exact eigenenergies show
a rigid shift $E_{nk_g\eta}-l_{0}^2/8l_{2}^2$.\cite{Gov04} 

For $\theta=\pi/2$ [Fig.~\ref{fig2}(b)] 
we find new energy degeneracies
entirely due to the presence of RIC
in addition to the anticrossings similar to the $\theta=0$ case.
For $0<\theta<\pi/2$ there is a local minimum
or maximum at $k_g$ depending on the size of $\alpha_2$ with
respect to $\alpha_1$ and $B$.
Our results, thus, demonstrate that including
intersubband coupling is essential in any theory of strong Rashba interaction
in quantum wires in the presence of magnetic fields and for $k$ values
near the subband gaps.

To see how sensitive the local maxima
are to the effect of RIC we calculate the second-order correction
in $\varH_{\alpha_2}$ to the energy spectrum at $\theta=0$ and then
find the dimensionless
effective mass at the extremum of the $n\eta$ branch,
$m/m^*_{n\eta}=\partial^2E_{nk\eta}/\partial (kl_{0})^2|_{k=k_g}$.
The sign of $m^*$ determines
the character of the extremum, electron-like ($m^*>0$)
or hole-like ($m^*<0$):
\begin{multline}\label{effectivemass}
\frac{m}{m^*_{n\eta}}
=1+\frac{\eta}{2}\left(\frac{l_Z}{l_{1}}\right)^2\\
\times\left[1-\frac{1}{2}\left(\frac{l_0}{l_{2}}\right)^2 
\frac{1+2n+\sqrt{2}\eta(l_0/l_Z)}{1-2(l_0/l_Z)^2}\right]\,.
\end{multline}
For $\alpha_2=0$ and large $\alpha_1$ ($l_{1}<l_Z/\sqrt{2}$),
$m^*_{n\eta}$ becomes negative for $\eta=-1$ subbands, leading
to a maximum in the energy spectrum, in agreement with our previous conclusions.
However, for strong enough RIC the
effective mass sign is reversed. We emphasize that this is an effect purely
due to the RIC and it is even stronger for increasing $n$, confirming
the numerical results of Fig.~\ref{fig2}(a).
Equation~(\ref{effectivemass}) is a perturbative result
and works well provided the confinement is the smallest length
scale ($l_0\ll l_Z,l_1,l_2$).
Only for very wide wires ($l_1,l_2,l_Z\ll l_0$)
we numerically recover maxima ($m^*_{n\eta}<0$).

\begin{figure}[t]
\centerline{\psfig{figure=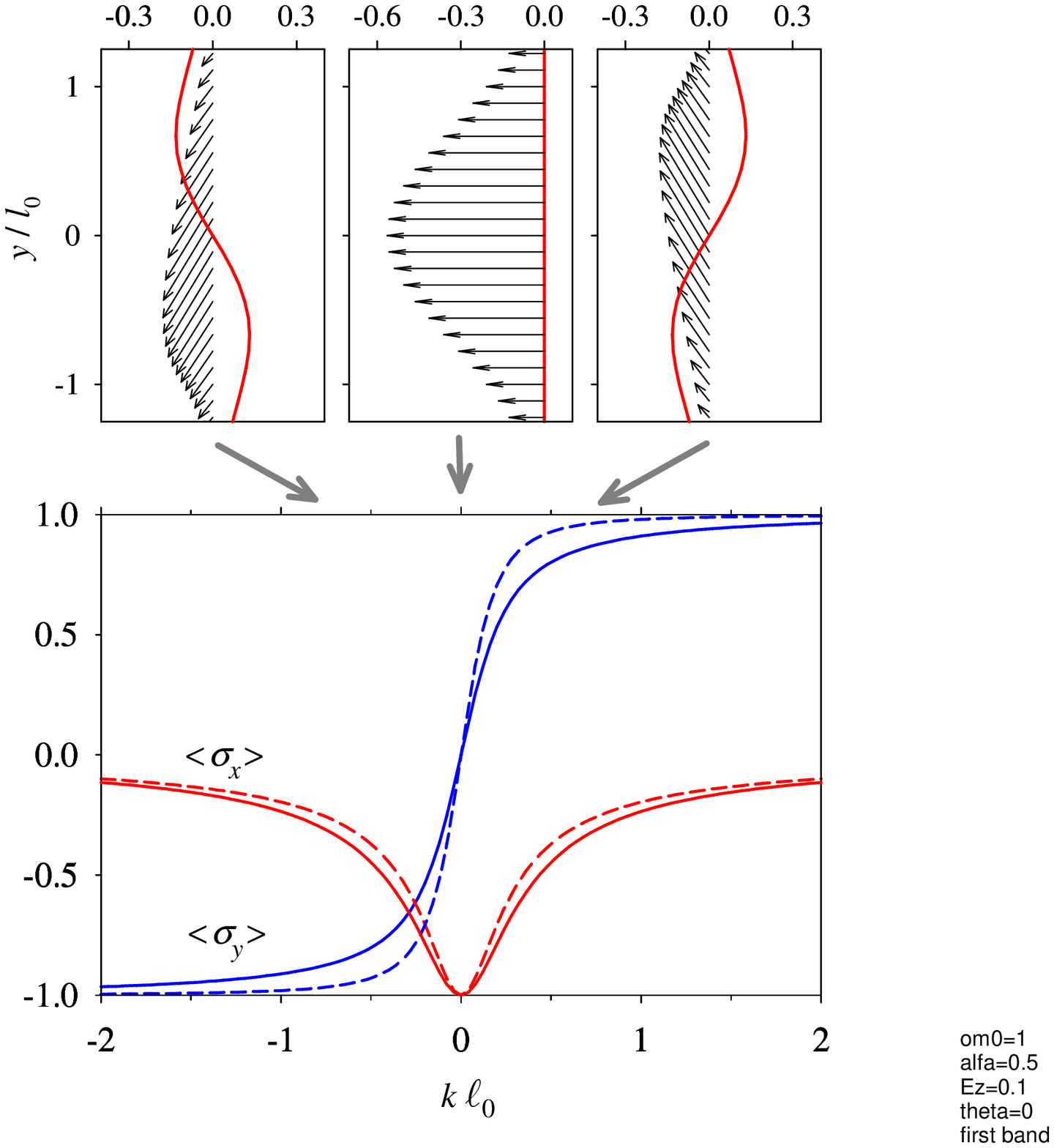,width=2.5in,clip=}}
\caption{(Color online) Lower plot: Dependence with $k$ of the spin 
expectation values in the lowest subband. Solid lines are obtained
including both Rashba terms while dashed lines are the results when
RIC is neglected.
We set $l_Z=2.2l_0$, $l_1=l_2=l_0$ and $\theta=0$. 
Upper panels display the spin texture 
for three selected propagation momenta and taking into account both Rashba 
terms. The $k l_0$ values (indicated by the thick arrows pointing on the $k$-axis) 
are -0.75, 0 and 0.75 for left, center and right upper plots, respectively. 
The vector plot shows the in-plane spin and the continuous line corresponds
to the $z$ component.} \label{fig3}
\end{figure}

\section{Spin orientation and magnetization}

In general, when both Rashba terms
are present spin is not a good quantum
number and we find spin textures, with the spin direction 
depending on $k$ and the wire transversal coordinate $y$. 
On the contrary, if RIC is neglected the states 
are proper eigenspinors even in the presence of
an in-plane magnetic field,\cite{Yur04} as seen from Eq.\ (\ref{states_s1}).
Therefore, the existence of clear spin textures is a
signature of the RIC term. 
The local spin components
for the lowest subband at three different propagation momenta
are shown in the upper plots of Fig.~\ref{fig3}.
While all in-plane spins are essentially collinear,   
a sizeable $z$-component, similar in magnitude to the in-plane one, 
precludes the definition of a common spin axis when $kl_0\ne0$
and thus shows the importance of RIC.
The local $z$-magnetization in real space
$\langle \sigma_{z}(y) \rangle$ is
antisymmetric in $y$, leading to a vanishing integrated $\langle
\sigma_z\rangle$,\cite{gov02} and giving rise to spin accumulations at the wire
edges which are reminiscent of the intrinsic spin Hall effect,\cite{Jai04} but
here the effect arises in a confined system.\cite{Usa05}.

\begin{figure}[t]
\centerline{\psfig{figure=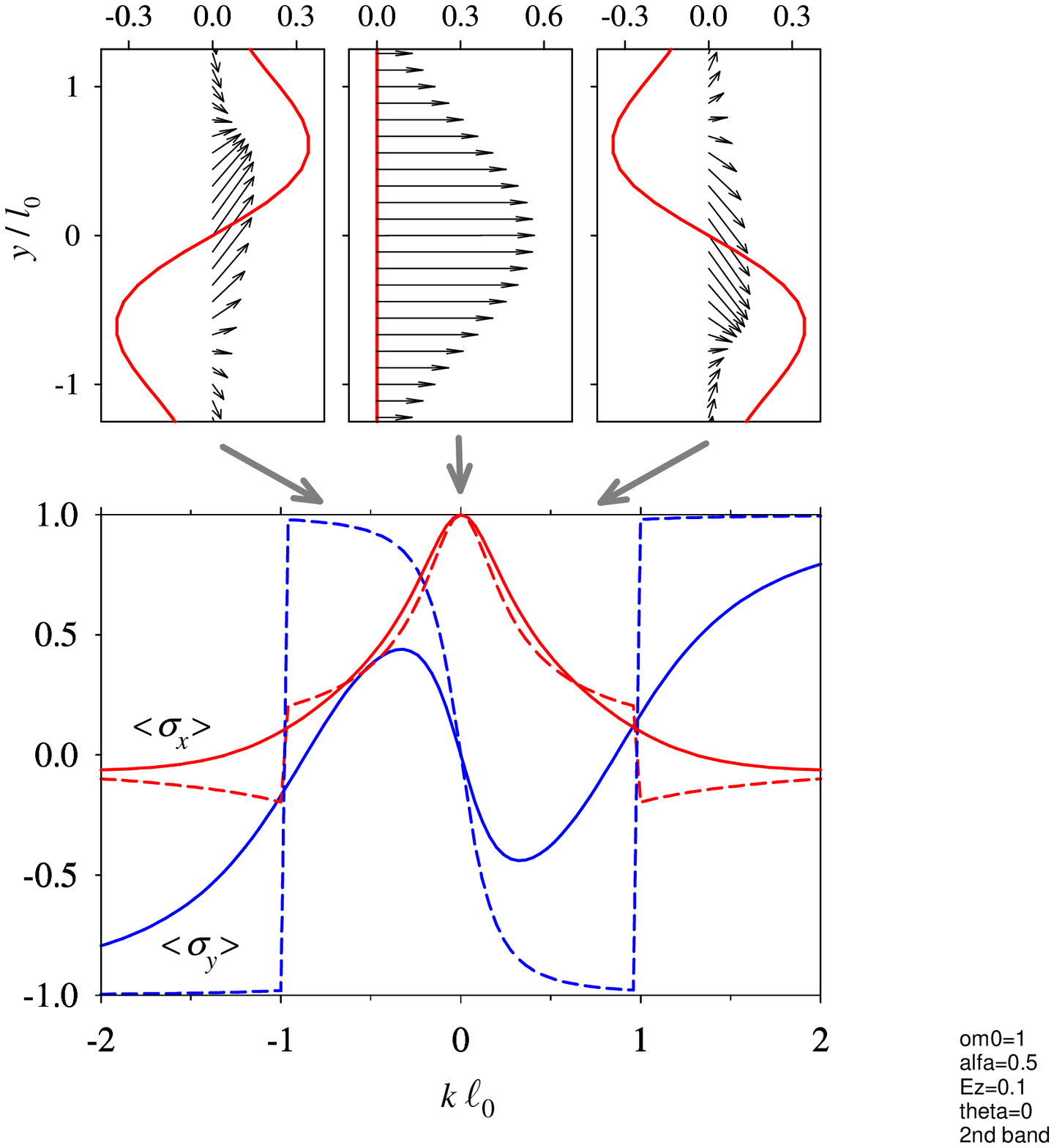,width=2.5in,clip=}}
\caption{(Color online) Same as Fig. 3 for the second subband.} 
\label{fig4}
\end{figure}

There are exceptional cases where the spin is well defined even
in the presence of both magnetic field and RIC. For $|k l_0|\gg 1$ one always
finds asymptotic eigenstates of $\sigma_y$,\cite{gov02} irrespectively of the
magnetic field orientation $\theta$, as seen in the lower plot of Fig.\ \ref{fig3}. For
states with large negative (positive) $k$ the spin is $\langle\sigma_y\rangle=-1(+1)$. 
Importantly, for $k=0$ (the minimum gap point for $\theta=0$)
we numerically find $\langle\sigma_x\rangle=\eta$, 
{\em regardless of the strengths of magnetic fields and intersubband coupling}.
This is clear from Eq.~(\ref{eq_exact}), which shows that
the spinors are eigenstates of $\sigma_x$ at the minimum gap momentum.
Therefore, there is always a given wavector $k_g=(l_1/l_Z^2)\sin\theta$,
satisfying $\Omega_{k_g}=0$, for which the propagation
direction is also the spin quantization axis.
Away from this special point, $\sigma_x$ tends to zero 
(see Fig.~\ref{fig3}) more slowly for stronger fields. 
Lower plot in Fig.~\ref{fig3} displays also the spin expectation values
$\langle\sigma_x\rangle $ and $\langle\sigma_y\rangle$ when RIC is
not taken into account (dashed lines). As we will see later,
the effect of RIC in higher subbands is more notable but
we emphasize that even in the lowest subband the effect of
Rashba intersubband coupling is not negligible due to the
observation of spin textures.

We plot in Fig.~\ref{fig4} the spatially dependent magnetization
and the spin expectation values for the second subband with
the same parameters as in Fig.~\ref{fig3}.
The effect of RIC is more pronounced than in the lowest subband
as seen from the spin textures, which contain a noncollinear distribution 
even for the in-plane spin.   
In addition the local $z$-magnetization is bigger 
than in the lower subband and exceeds the horizontal components,
leading to large spin accummulations at the wire edges. For $k=0$ we 
again find that all local spins are properly aligned in the direction of 
transport. 
Remarkably,
the spin expectation values $\langle\sigma_x\rangle$ and
$\langle\sigma_y\rangle$ change drastically when the Rashba interaction is
fully included as shown in lower Fig.~\ref{fig4}. When RIC is absent,
both expectation values display abrupt jumps due to the subband crossings. As
we pointed out earlier, RIC avoids these crossings and this is reflected
in a much smoother behavior of $\langle\sigma_x\rangle$ and
$\langle\sigma_y\rangle$. 

Opposite sign  accumulations of the $z$-magnetization at the wire edges 
are already found in the $B=0$ case.\cite{Gov04,gov02}
Differences with respect to the zero field case, however,
can be summarized in that the magnetic field 
causes nonzero $x$-magnetization across the wire for wavevectors around
the minimum gap point. The tilted distributions of in-plane 
magnetizations in Fig.\ 3 are due entirely to the $x$-component of the
field. Additionally, the texturing of the horizontal magnetization 
of the second subband (Fig.\ 4) is due to the combination
of $B$ and RIC. The magnetic field also shifts the minimum gap
momentum $k_g$ from its zero value at vanishing $B$, for which an 
exact solution was 
obtained in the preceding section.

\begin{figure}[t]
\centerline{\psfig{figure=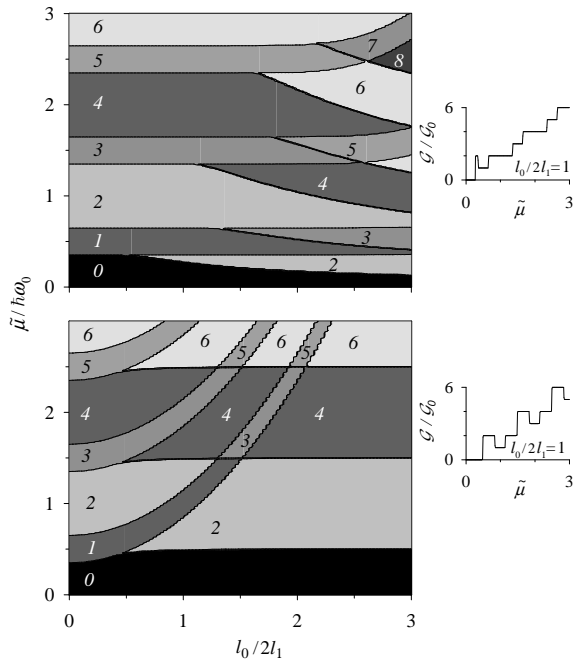,width=3.in,clip=}}
\caption{Variation of the conductance with Rashba intensity 
and the Fermi energy $\mu$ in the full Rashba model ($l_2=l_1$, upper panel) 
and neglecting RIC ($l_2=\infty$, lower panel). We set $l_Z=1.8l_0$
and parallel magnetic field $\theta=0$.
A rigid shift with Rashba intensity 
has been taken into account defining $\tilde\mu/\hbar\omega_0=\mu/\hbar\omega_0+l_0^2/8l_1^2$.
The numbers in the plateaus give the conductance in units of the conductance 
quantum. Right small plots show vertical cuts of the corresponding left figures for 
the given values of $l_0/2l_1$.}
\label{fig5}
\end{figure}

\section{Transport properties}

The linear conductance $\mathcal{G}$ at zero temperature is 
directly related to the wire subband structure through the formula,
\cite{Mor99,Yur04} 
\begin{equation}\label{conductance}
\mathcal{G}=\frac{e^2}{h} \sum_{i j}{ \Theta(\mu-\xi^{(j)}_{i})
\, {\rm sgn}(m^{(j)*}_{i})}\;,
\end{equation}
which accounts for the number of occupied subbands propagating in a
given direction and the conductance quantum per spin component, 
${\mathcal G}_0=e^2/h$.
In Eq.~(\ref{conductance}),
$\mu$ is the Fermi energy while $\xi^{(j)}_{i}$ and $m^{(j)*}_{i}$ denote,
respectively, the energy
and effective mass at the $j$-th local extremum in the $i$-th subband. 
We have also defined $\Theta(x)$ and ${\rm sgn}(x)$ as the step and the 
sign functions. 
For subbands containing a single minimum  
$\mathcal{G}$ only attains
$e^2/h$ steps when $\mu$ is increased.
In the case of more complicated subbands 
it is easy to see from Eq.~(\ref{conductance}) that the 
linear conductance step is doubled whenever the
Fermi energy exceeds two local minima but is below the energy of the maximum 
lying in between them. Once $\mu$ exceeds the energy of the maximum, the conductance
is decreased by $e^2/h$.\cite{Yur04,nes04} 
As discussed above the RIC can convert subband maxima into minima and, therefore,
$\mathcal{G}$ is generally reduced when both terms are included.

For $\theta=0$ we plot in Fig.~\ref{fig5} the linear conductance for the cases
with RIC, $l_{2}=l_{1}$ (upper panel), and without it, $l_{2}=\infty$ (lower
panel). When RIC is not included, $\mathcal{G}$ alternates,
with increasing $\mu$, steps of $+2e^2/h$
with downward jumps of $-e^2/h$ due to the presence of maxima in the 
lower subbands.\cite{Yur04}
When RIC is included
two major modifications are apparent: a) Stronger Rashba couplings are needed to
observe the alternate steps of $+2e^2/h$ and $-e^2/h$, and
b) in any case this anomalous pattern of steps vanishes when increasing the Fermi
energy. Particularly, as shown by the insets, when $l_0/2l_1=1$ the sequence of 
conductance steps in units of ${\cal G}_0$ is $+2,-1,+1,+2,-1,+1,\dots$  
when RIC is neglected, and $+2,-1,+1,+1,+1,\dots$ when it is included.
Therefore, Fig.~\ref{fig5} proves that to observe
the modifications of the linear conductance steps due to the Rashba
interaction it is essential to have a relatively low Fermi energy
or, equivalently, a rather small number of propagating modes.

\begin{figure}[t]
\centerline{\psfig{figure=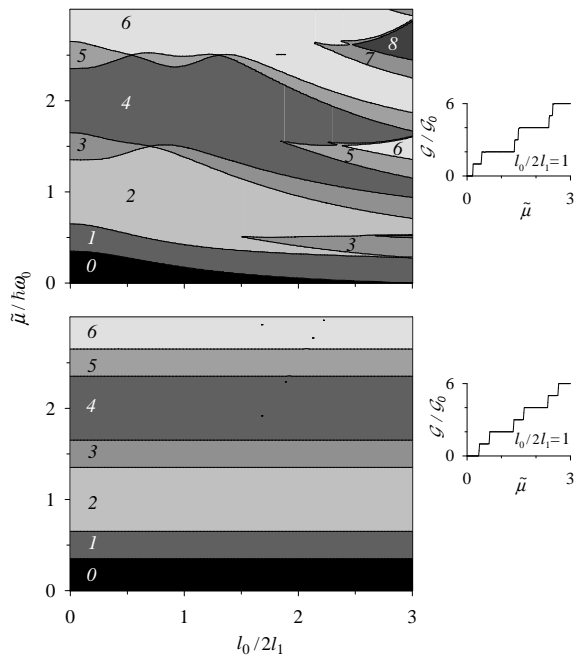,width=3.in,clip=}}
\caption{Same as Fig. 5 for perpendicular field ($\theta=\pi/2$).}
\label{fig6}
\end{figure}

Figure~\ref{fig6} shows the case $\theta=\pi/2$. In lower Fig.~\ref{fig6},
when $l_2=\infty$, the conductance consists of regular jumps of one
conductance quantum. From Fig.~\ref{fig2}(b) we determine that
the width of the odd conductance plateaus corresponds to
the Zeeman energy whereas the width of the even plateaus amounts
to $1-(l_0/l_Z)^2$ (in units of $\hbar \omega_0$).
This physical scenario is strongly modified when RIC is taken
into account, as seen in the upper Fig.~\ref{fig6} where
the conductance shows a much richer behavior.
It is remarkable the occurrence of
{\em restoration of the spin degeneracy} for special values of the Rashba
coupling and magnetic field where the third and fifth conductance plateaus
collapse. In general, the tendency is to suppress regions for \emph{odd}
values of $\mathcal{G}$. Thus, $\mathcal{G}$ presents wide (narrow) plateaus
for even (odd) values of the conductance quantum but their
widths are not trivially related to either the Zeeman energy
or the confinement energy alone. Besides, at large Rashba strengths we
also find decreasing jumps in the conductance.

\section{Conclusions}

We have discussed the spectrum, spin orientation and linear
transport of ballistic quantum wires in the presence of Rashba
interaction and in-plane magnetic fields at an angle $\theta$
with the propagation direction. Analytical and 
numerical results demonstrating the crucial importance
of the Rashba intersubband coupling for all these properties
and for $k$ values near the subband gaps
have been presented. At strong spin-orbit coupling the modifications 
in the energy subbands 
include the appearance of sizeable shifts, anticrossings as well as 
big reductions of the subband maxima. The changes in subband 
structure lead 
to precise predictions for measurements of the linear 
transport properties, such as a severe reduction of the anomalous 
conductance steps when $\theta=0$ and a nontrivial dependence 
of the steps with the Rashba intensity for $\theta=\pi/2$, with 
collapsing points for the odd plateaus.

Spin textures in the local magnetization are obtained 
only when the Rashba intersubband coupling is
taken into account. Opposite-sign accumulations of the spin $z$-magnetization 
at the wire edges are strongly dependent
on the $k$-value and subband index.
For subbands above the lower one we have shown that 
the horizontal magnetization is also textured. At a particular momentum 
(the minimum-gap momentum), depending
on magnetic field strength, orientation, and Rashba intensity, 
the analytical solution having the spin along the transport
direction has been obtained. We believe these results are relevant for 
the future design of spin transistors and spin filters.
 
\section*{Acknowledgements}
This research was supported by the Spanish grants
PRIB-2004-9765 (Govern de les Illes Balears),  
BFM2002-03241 (MEC), and the ``Ram\'on y Cajal'' program.


\begin{thebibliography}{90}
\bibitem{Ras60}
E.I.~Rashba, Fiz. Tverd. Tela (Leningrad)  {\bf 2}, 1224 (1960).
[Sov.\ Phys.\ Solid State {\bf 2}, 1109 (1960)].
\bibitem{Nit97}
J.~Nitta, T.~Akazaki, H.~Takayanagi, and T. Enoki, Phys. Rev. Lett {\bf 78}, 1335 (1997).
\bibitem{Eng97}
G.~Engels, J.~Lange, Th. Sch\"apers, and H. L\"uth, Phys. Rev. B {\bf 55}, R1958 (1997).
\bibitem{Scha04}
Th. Sch\"apers, J.~Knobbe and V.A~Guzenko Phys. Rev. B {\bf 69}, 235323 (2004).
\bibitem{Dat90}
S.~Datta and B.~Das, Appl. Phys. Lett.  {\bf 56}, 665 (1990).
\bibitem{stre03}
P. Streda and P. Seba
Phys. Rev. Lett. {\bf 90}, 256601 (2003).
\bibitem{egu02}
J.C. Egues, G. Burkard, and D. Loss 
Phys. Rev. Lett. {\bf 89}, 176401 (2003).
\bibitem{nes04}
J.A. Nesteroff, Yu.V. Pershin, and V. Privman,
Phys. Rev. Lett. 93, 126601 (2004).
\bibitem{Mor99}
A.V.~Moroz and C.H.W.~Barnes, Phys. Rev. B {\bf 60}, 14272 (1999).
\bibitem{Mir01}
F.~Mireles and G. Kirczenow, Phys. Rev. B {\bf 64}, 024426 (2001).
\bibitem{Yur04}
Yu.V. Pershin, J.A. Nesteroff, and V. Privman,
Phys. Rev. B {\bf 69}, 121306 (2004).
\bibitem{cah04}
M. Cahay and S. Bandyopadhyay,
Phys. Rev. B {\bf 69}, 045303 (2004).
\bibitem{per05}
R.G. Pereira and E. Miranda,
Phys. Rev. B {\bf 71}, 085318 (2005).
\bibitem{Byc90} 
Yu.\ A. Bychkov, V. I. Mel'nikov, E. I. Rashba,
Zh.\ Eksp.\ Teor.\ Fiz.\ {\bf 98}, 717 (1990)
[Sov.\ Phys.\  JETP {\bf 71}, 401 (1990)].
\bibitem{Deb04}
S.~Debald and B. Kramer, Phys. Rev. B {\bf 71}, 115322 (2005).
\bibitem{notefoot}
If $g<0$, the global sign of the Zeeman term should be reversed.
\bibitem{Che99}
G.-H. Chen and M.~E.Raikh, Phys. Rev. B. {\bf 60}, 4826 (1999).
\bibitem{Val03} M. Val\'{\i}n-Rodr\'{\i}guez, A. Puente, Ll.\ Serra, 
Eur.\ Phys.\ J. B {\bf 34}, 359 (2003). 
\bibitem{Gov04} This result generalizes the $B=0$ case recently discussed by
  M.~Governale and U.~Z\"ulicke, Solid State Comm.  {\bf 131}, 581 (2004).
\bibitem{gov02}
M. Governale and U. Z\"ulicke, 
Phys. Rev. B {\bf 66}, 073311 (2002).
\bibitem{Jai04}
J.~Sinova, D.~Culcer, Q.~Niu, N.A.~Sinitsyn, T.~Jungwirth, and A.H.~MacDonald,  Phys. Rev. Lett. {\bf 92}, 126603 (2004).
\bibitem{Usa05}
G.~Usaj and C.A.~Balseiro, cond-mat/0405065 (unpublished).
\end{thebibliography}
\end{document}